\documentstyle[jetsu1_epsf]{laa}        % laa A&A with Figs.
\def\picture#1#2{\picturePS{#1}{#2}}    % Include PS figure only
\begin{document}                        % Do not leave blanks in here
%% __________________________________________________________________________|
\newcommand{\araa}{{ARA{\&}A~}} % Annual Review of Astronomy and Astrophysics
\newcommand{\aj}{{AJ~}}         % Astronomical Journal
\newcommand{\ana}{{A{\&}A~}}    % Astronomy and Astrophysics 
\newcommand{\anas}{{A{\&}AS~}}  % Astronomy and Astrophysics Supplement Series
\newcommand{\aar}{{A{\&}AR~}}   % Astronomy and Astrophysics Review
\newcommand{\apj}{{ApJ~}}       % Astrophysical Journal 
\newcommand{\apjs}{{ApJS~}}     % Astrophysical Journal Supplement Series
\newcommand{\apss}{{Ap{\&}SS~}} % Astrophysics and Space Science
\newcommand{\mnras}{{MNRAS~}}   % Monthly Notices of Royal Astronomical Society
\newcommand{\pasp}{{PASP~}}     % Publications of the Astronomical Society Pac.
\newcommand{\paspc}{{PASPC~}}   % Publ. Astron. Soc. Pac. Conf. proceedings
\newcommand{\nat}{{Nat~}}       % Nature
%% __________________________________________________________________________|
\newcommand{\solp}{{Solar Phys.~}}
\newcommand{\csss}{{Cool Stars, Stellar Systems, and the Sun~}}
\newcommand{\iappp}{{IAPPP Commun.~}}
\newcommand{\ibvs}{{Inf. Bull. Variable Stars~}}
\newcommand{\aspas}{{ Astrophys. Space Sci.~}}
\newcommand{\ika}{{ Izv. Krymsk. Astrofiz. Obs.~}}
%% __________________________________________________________________________|
% Samples
\newcommand{\crione}{{\sc{c}}$_1$}
\newcommand{\critwo}{{\sc{c}}$_2$}
\newcommand{\crithree}{{\sc{c}}$_3$}
\newcommand{\crifour}{{\sc{c}}$_4$}
\newcommand{\crifive}{{\sc{c}}$_5$}
\newcommand{\crisix}{{\sc{c}}$_6$}
\newcommand{\criseven}{{\sc{c}}$_7$}
\newcommand{\crieight}{{\sc{c}}$_8$}
% weights
\newcommand{\wwt}{$w_{\rm t}$}
\newcommand{\wwd}{$w_{\rm D}$}
% minor
\newcommand{\xs}{ \! \cdot \! }
\newcommand{\lj}{{\tiny \: \pm \: }}
\newcommand{\human}{{$\sim 0$}}
\newcommand{\signal}{``human--signal''}
\newcommand{\never}{0}
\newcommand{\reject}{~$\bullet$}
\newcommand{\nostat}{$\ast$}
\newcommand{\mlow}{{$\dagger$}}
\newcommand{\mhigh}{{$\ddagger$}}
% Methods
\newcommand{\SD}{SD--method}% _______________________________________________|
%% Do not leave blanks in here
\newcommand{\WSD}{WSD--method}
\newcommand{\K}{K--method}
\newcommand{\WK}{WK--method}
\newcommand{\TWOK}{K$^{2}$--method}
\newcommand{\SHORTSD}{SD--}
\newcommand{\SHORTWSD}{WSD--}
\newcommand{\SHORTK}{K--}
\newcommand{\SHORTWK}{WK--}
%% __________________________________________________________________________|
   \thesaurus{07                 % Solar system:                    % OK 
             (07.05.1,           % Solar system: Earth              % OK 
              07.03.1,           % Solar system: Comets             % OK 
              03.13.6,           % Methods: Statistical             % OK
              10.19.1)           % Galaxy: solar neighbourhood      % OK
                        }        %
% ___________________________________________________________________________|
\title{The ``human'' statistics of terrestrial impact cratering rate\thanks{
Table 1 is only available in electronic form: see editorial in A\&A
1992, Vol 266, page E1}} 
% ___________________________________________________________________________|
\author{   L. Jetsu}                        

\offprints{L. Jetsu (jetsu@nordita.dk)}
 
\institute{ NORDITA, Blegdamsvej 17, Copenhagen 2100, 
            Denmark }                      

\date{Received; accepted}

\maketitle

\begin{abstract}
% __________________________________ Do not leave a blank line here! _______|
The most significant periodicities in the terrestrial impact
crater record are due to the \signal: 
the bias of assigning integer values for the crater ages.
This bias seems to have eluded the proponents and opponents 
of real periodicity in the occurrence of these events, 
as well as the theorists searching for an extraterrestrial explanation
for such periodicity.
The \signal ~should be seriously considered by scientists in astronomy,
geology and paleontology when searching for a connection
between terrestrial major comet or asteroid impacts and mass extinctions
of species.

\end{abstract}
% __________________________________________________________________________|

\section{Introduction}

An outstanding series of papers appeared in 1984 when
a 28.4~Myr cycle was detected in the terrestrial impact crater record
(Alvarez \& Muller 1984, Davis et al. 1984, Whitmire \& Jackson 1984). 
This value was 
close to the 26~Myr cycle discovered in the geological record of 
major mass extinctions of species (Raup \& Sepkoski 1984).
The fascinating idea of periodic comet impacts 
causing ecological catastrophies emerged
(Alvarez \& Muller 1984, Davis et al. 1984).
It was suggested that an unseen solar companion (Nemesis) 
might induce gravitational disturbances to the Oort comet cloud 
triggering periodic cometary showers 
(Davis et al. 1984, Whitmire \& Jackson 1984). Other astronomical models have
been proposed later to account for the above periodicities, 
the ``galactic carrousel'' being perhaps the most 
widely accepted model (e.g. the review by
Rampino \& Haggerty 1996, and references).
The main idea of the ``galactic carrousel'' model is that the
Oort comet cloud is periodically perturbed by galactic tides
as the Solar System revolves around the centre of the
Milky Way galaxy.

We will show that only one extremely 
significant regularity exists in the impact crater record: the \signal.

\section{Data}

We chose $n\!=\!82$ impact craters with a diameter $D_{\rm i}$ [km] 
and an age $t_{\rm i}$ [Myr], which had an error ($\sigma_{\rm t_i}$)
in the database maintained by the Geological Survey of Canada.\footnote{ 
//gdcinfo.agg.emr.ca/cb-bin/crater/crater\_table?e} 
These data are {\it only} published in electronic form in our Table 1.
Eight simultaneous events were combined ($\Rightarrow \! n \! = \! 74$)
with the relations
$\sigma_{\rm 1,2}={\rm min}[\sigma_{\rm t_1},\sigma_{\rm t_2}]/\sqrt{2}$
and
$D_{1,2}=(D_1^2+D_2^2)^{1/2}$, 
where $\sigma_{\rm t_1}$, $\sigma_{\rm t_2}$, $D_1$ and $D_2$
refer to the individual events.  
The geographical coordinates of these pairs imply an occurrence of
a double impact, except for one pair ($i=37$).
Combining probable double impacts leaves our result unchanged,
but provides better statistics. Six subsamples were selected from
Table 1 ($n\!=\!74$):
\begin{itemize}

\item[\crione:] $5 \le t$

\item[\critwo:] $t \le 250$, $\sigma_{\rm t} \le 20$, $D \ge 5$

\item[\crithree:] $5\le t \le 300$, $\sigma_{\rm t} \le 20$

\item[\crifour:] \crione, $t$ is not a multiple of 5

\item[\crifive:] \critwo, $t$ is not a multiple of 5

\item[\crisix:] \crithree, $t$ is not a multiple of 5

\end{itemize}
{\noindent 
Criteria similar to \critwo ~and \crithree ~have been applied 
earlier (Grieve \& Pesonen 1996, Matsumoto \& Kubotani 1996). 
We also analysed the sample (\criseven: $n\!=\!13$),
where the 28.4~Myr cycle was originally detected
(Alvarez \& Muller 1984), and
one sample of mass extinctions of species (\crieight: n=8),
recently compared to the impact crater record
(Matsumoto \& Kubotani 1996).
Two sets of weights were derived: }
\begin{eqnarray}
w_{\rm t_i}     =  A_{\rm t}~ \sigma_{\rm t_i}^{-2} {\rm ~and~}
w_{\rm D,i}     =  A_{\rm D}~ D_{\rm i},            \nonumber
\end{eqnarray}
{\noindent where the constants
$A_{\rm t}  \! = \! n  /  [\sum_{\rm i=1}^{\rm n} \sigma_{\rm i}^{-2}$] and
$A_{\rm D}  \! = \! n  /  [\sum_{\rm i=1}^{\rm n} D_{\rm i}$] ensure that 
$\sum_{\rm i}^{\rm n} w_{\rm t_i} \! = \!
 \sum_{\rm i}^{\rm n} w_{\rm D_i} \! = \! = \!n$.
The two largest ($w_{\rm max,1}$, $w_{\rm max,2}$) and the smallest 
($w_{\rm min}$) weights, 
the ratio $W_{\rm R}\!=\!w_{\rm max,1}/w_{\rm min}$ and the average 
$s\!=\!(w_{\rm max,1}\!+\!w_{\rm max,2})/2$ for \crione, ..., \criseven
~are given in Table 2 (No $\sigma_{\rm t_i}$ were available for \crieight).}

\addtocounter{table}{+1}               
\begin{table} % _____________________________________________________________|
\caption[ ]{The weights \wwt ~and \wwd ~of \crione, ..., \criseven:
the two largest ($w_{\rm max,1}$, $w_{\rm max,2}$) and the smallest
($w_{\rm min}$) weights, the ratio $W_{\rm R}\!=\!w_{\rm max,1}/w_{\rm min}$,
the average $s\!=\!(w_{\rm max,1}+w_{\rm max,2})/2$, and the breakdown
parameter $R(s)$ (Jetsu 1996: Eq. 5). }
\addtolength{\tabcolsep}{-0.12cm}  
\renewcommand{\arraystretch}{0.92}
\begin{tabular}{lccrrcrr}
\hline
~~~~$w$                              & 
$n$                                  &
$w_{\rm min}$                        &
$w_{\rm max,1}$                      &
$w_{\rm max,2}$                      &
$W_{\rm R}$                          & 
$s$~~~                               &
$R(s)$                               \\
\hline % _______________________________ KUVAUS.TEX ________________________%|
{\crione}(\wwt)     & 
61  & $3.15\xs10^{-6}$  & 50.41 & 6.43  & $1.60\xs10^{7}$  & 28.42 & 3.57 \\
{\crione}(\wwd)     & 
61  & $4.44\xs10^{-2}$  & 8.32  & 6.93  & 187.50           & 7.63  & 0.10 \\
\hline
{\critwo}(\wwt)     & 
34  & $2.75\xs10^{-5}$  & 27.54 & 4.41  & $1.00\xs10^{6}$  & 15.97 & 4.21 \\
{\critwo}(\wwd)     & 
34  & 0.15              & 4.66  & 2.74  & 30.91            & 3.70  & 0.10 \\
\hline
{\crithree}(\wwt)     & 
35  & $1.81\xs10^{-4}$  & 29.00 & 3.70  & $1.60\xs10^{5}$  & 16.35 & 3.94 \\
{\crithree}(\wwd)     & 
35  & $6.99\xs10^{-2}$  & 4.75  & 2.80  & 68.00            & 3.77  & 0.10 \\
\hline
{\crifour}(\wwt)     & 
27  & $1.16\xs10^{-4}$  & 22.44 & 2.86  & $1.94\xs10^{5}$  & 12.65 & 4.14 \\
{\crifour}(\wwd)     & 
27  & $6.47\xs10^{-2}$  & 6.26  & 3.55  & 96.77            & 4.91  & 0.23 \\
\hline
{\crifive}(\wwt)     & 
25  & $2.03\xs10^{-5}$  & 20.26 & 3.24  & $1.00\xs10^{6}$  & 11.75 & 4.30 \\
{\crifive}(\wwd)     & 
25  & 0.15              & 4.66  & 2.74  & 30.91            & 3.70  & 0.17 \\
\hline
{\crisix}(\wwt)     & 
23  & $1.20\xs10^{-4}$  & 19.15 & 2.44  & $1.60\xs10^{5}$  & 10.80 & 4.23 \\
{\crisix}(\wwd)     & 
23  & $8.17\xs10^{-2}$  & 4.48  & 2.63  & 54.84            & 3.56  & 0.19 \\
\hline
{\criseven}(\wwt)     & 
13  & $1.09\xs10^{-2}$  & 8.90  & 1.09  & 816.33           & 4.99  & 1.41 \\
{\criseven}(\wwd)     & 
13  & 0.15              & 2.94  & 2.36  & 20.00            & 2.65  & 0.33 \\
\hline % ___________________________________________________________________%|
\end{tabular}
\end{table}
\addtolength{\tabcolsep}{0.12cm}  
\renewcommand{\arraystretch}{1.00}
%% __________________________________________________________________________|

%% __________________________________________________________________________|
\begin{figure*}
\picture{3.2cm}{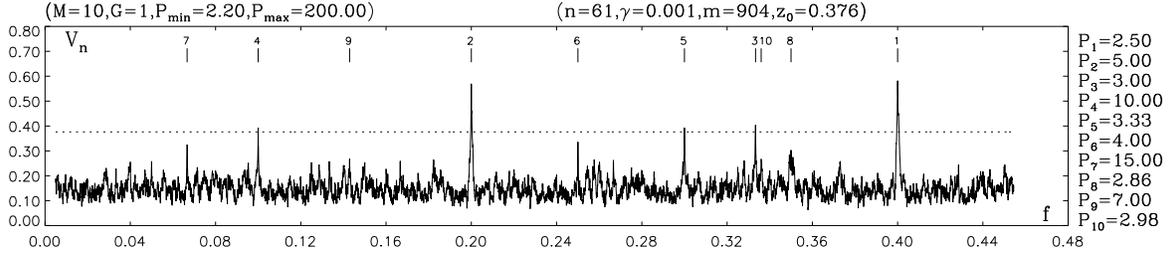}
\caption[]{The ten best period candidates detected with the \K ~for
\crione ~with $n\!=\!61$: the overfilling factor (Jetsu \& Pelt 1996: Eq. 12) 
is $[\Delta\phi]^{-1}\!=\!GM=10$ for this test between $P_{\rm min}\!=\!2.2$
and $P_{\rm max}\!=\!200$. The statistics for $m\!=\!904$ independent
frequencies yield the level $P(V_{\rm n} \ge z_0) \!=\! \gamma \!=\!0.001$
outlined with a horizontal line. The numbers above the 
short vertical lines denote the locations of $P_1, ..., P_{10}$. }
\end{figure*}
%% __________________________________________________________________________|

%% __________________________________________________________________________|
\begin{figure*}
\picture{3.2cm}{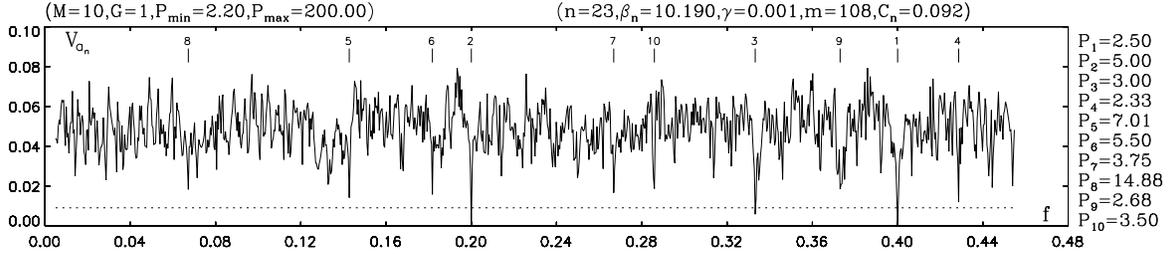}
\caption[ ]{The ten best period candidates detected with the \WSD ~for
\crisix(\wwd) with $n\!=\!23$: the level 
$P(V_{\rm a_n} \le C_{\rm n}/\beta_{\rm n})\!=\!\gamma \!=\!0.001$ for
$m\!=108$ is
outlined with a horizontal line ($P_{\rm min}$, $P_{\rm max}$, $G$ and
$M$ as in Fig. 1).}
\end{figure*}
%% __________________________________________________________________________|

\begin{table*} % _____________________________________________________________|
\caption[ ]{The five best periods detected in \crione, ..., \crieight ~with
the \SHORTK, \SHORTWK, \SHORTSD ~and {\WSD}s: The estimate of the number
of independent frequencies is $m$ for a sample of a size $n$.
The symbols \mhigh ~and \mlow ~denote the cases $m\!>\!m'$ and $m\!<m\!'$.
The critical levels for each period $P$ are 
$Q_{\rm K}$ $Q_{\rm WK}$ $Q_{\rm SD}$ and $Q_{\rm WSD}$. The rejection of
$H_0$ with $\gamma \!=\!0.001$ is indicated by \reject. Finally, the
cases when an application of some particular method would not yield reliable
results are denoted by \nostat}
\addtolength{\tabcolsep}{0.12cm}  
\renewcommand{\arraystretch}{0.95}
\begin{tabular}{|c|c|c|c|c|}
\hline % ___ Written by hand from the large document _______________________%|
                    & 
\K                  &
\WK                 &
\SD                 &
\WSD                \\
                    &
  no weights        &
  \wwd              &
  no weights        &
  \wwd              \\
                    &
$P$ ($Q_{\rm K}$)   &
$P$ ($Q_{\rm WK}$)  &
$P$ ($Q_{\rm SD}$)  &
$P$ ($Q_{\rm WSD}$) \\
\hline % ___ Written by hand from the large document _______________________%|
\crione   &  2.50    (\human\mhigh)\reject          
          &  2.50    (\human\mhigh)\reject  
          & 20.00    (\never\mhigh)\reject          
          &  2.22... (\never\mhigh)\reject \\
$n=61$    &  5.00    ($2.0\xs10^{-13}$\mhigh)\reject          
          &  3.00    ($1.0\xs10^{-13}$\mlow)\reject
          &  2.66... (\never\mhigh)\reject          
          &  5.00    (\never\mhigh)\reject \\
$m=904$   &  3.00    ($3.9\xs10^{-6}$\mhigh)\reject 
          & 5.00     ($1.2\xs10^{-8}$\mlow)\reject
          &  5.00    (\never\mhigh)\reject          
          &  2.66... (\never\mhigh)\reject  \\
          & 10.00    ($2.1\xs10^{-4}$\mhigh)\reject 
          & 15.00    ($5.0\xs10^{-8}$\mlow)\reject
          &  8.00    (\never\mhigh)\reject          
          &  4.00    (\never\mhigh)\reject   \\
          &  3.33    ($2.1\xs10^{-4}$\mhigh)\reject 
          &  9.97    ($4.1\xs10^{-7}$\mlow)\reject
          &  2.50    (\never\mhigh)\reject          
          & 20.00    (\never\mhigh)\reject \\
\hline % ___________________________________________________________________%|
\critwo   &  3.00    (0.0078)                   
          &  2.50    ($2.6\xs10^{-5}$)\reject
          &  5.00    ($8.9\xs10^{-10}$)\reject  
          &  5.00    ($8.9\xs10^{-10}$)\reject  \\
$n=34$    &  2.50    (0.11)                     
          &  7.76    ($6.6\xs10^{-4}$)\reject
          &  2.50    ($8.9\xs10^{-10}$)\reject  
          &  2.50    ($8.9\xs10^{-10}$)\reject  \\
$m=110$   & 11.67    (0.58)                     
          &  3.00    ($7.4\xs10^{-4}$)\reject
          &  3.00    ($1.6\xs10^{-9}$)\reject   
          &  3.00    ($1.9\xs10^{-9}$)\reject   \\
          & 17.50    (0.62)                     
          &  9.95    (0.0037)
          &  2.25    ($4.0\xs10^{-9}$)\reject   
          &  6.00    ($6.7\xs10^{-9}$)\reject   \\
          &  7.00    (0.73)                     
          & 12.56    (0.0046)
          &  6.00    ($5.0\xs10^{-9}$)\reject   
          &  4.00    ($7.3\xs10^{-9}$)\reject   \\
\hline % ___________________________________________________________________%|
\crithree &  2.50    (0.0023)                   
          &  2.50    ($1.4\xs10^{-6}$)\reject
          &  5.00    ($4.2\xs10^{-10}$)\reject  
          &  2.50    ($4.2\xs10^{-10}$)\reject  \\
$n=35$    &  3.00    (0.0053)                   
          &  3.00    ($1.3\xs10^{-4}$)\reject
          &  2.66... ($4.2\xs10^{-10}$)\reject  
          &  8.00    ($4.2\xs10^{-10}$)\reject  \\
$m=127$   &  5.00    (0.18)                     
          &  7.76    ($1.5\xs10^{-4}$)\reject
          &  2.50    ($4.2\xs10^{-10}$)\reject  
          &  5.00    ($4.2\xs10^{-10}$)\reject  \\
          &  2.86    (0.58)                     
          &  12.56   (0.0016)
          &  4.00    ($4.2\xs10^{-10}$)\reject  
          &  4.00    ($4.2\xs10^{-10}$)\reject  \\
          & 11.62    (0.80)                     
          &  9.95    (0.0038)
          & 10.00    ($4.2\xs10^{-10}$)\reject  
          &  2.66..  ($4.2\xs10^{-10}$)\reject  \\
\hline % ___________________________________________________________________%|
\crifour  &  2.50    ($0.33$\mhigh)                  
          &  2.50    ($2.3\xs10^{-4}$\mlow)\reject
          &  4.00    ($3.1\xs10^{-6}$\mhigh)\reject  
          &  4.00    ($3.1\xs10^{-6}$\mhigh)\reject \\
$n=27$    & 13.20    ($0.83$\mhigh)                  
          &  2.98    ($5.3\xs10^{-4}$\mlow)\reject
          &  2.50    ($3.1\xs10^{-6}$\mhigh)\reject  
          &  2.50    ($3.1\xs10^{-6}$\mhigh)\reject \\
$m=904$   &  3.00    ($0.88$\mhigh)                  
          &  2.70    ($6.3\xs10^{-4}$\mlow)\reject
          &  5.00    ($3.1\xs10^{-6}$\mhigh)\reject  
          &  5.00    ($3.1\xs10^{-6}$\mhigh)\reject \\
          &  2.98    ($0.96$\mhigh)                  
          &  7.77    ($7.1\xs10^{-4}$\mlow)\reject
          & 10.00    ($3.1\xs10^{-6}$\mhigh)\reject  
          & 10.00    ($3.1\xs10^{-6}$\mhigh)\reject \\
          &  5.50    ($0.97$\mhigh)                  
          & 30.00    ($0.0016$\mlow)
          &  3.33... ($3.1\xs10^{-6}$\mhigh)\reject  
          &  3.33... ($3.1\xs10^{-6}$\mhigh)\reject \\ 
\hline % ___________________________________________________________________%|
\crifive  &  2.50    (0.17)                     
          &  7.77    (0.0011)           
          &  2.50    ($2.0\xs10^{-6}$)\reject   
          &  2.50    ($2.0\xs10^{-6}$)\reject  \\
$n=25$    & 17.43    (0.31)                     
          &  2.98    (0.0011)           
          &  5.00    ($2.0\xs10^{-6}$)\reject   
          &  5.00    ($2.0\xs10^{-6}$)\reject  \\
$m=110$   &  7.50    (0.49)                     
          &  3.71    (0.0033)           
          &  3.00    ($3.7\xs10^{-6}$)\reject   
          &  3.00    ($4.9\xs10^{-6}$)\reject  \\
          &  3.00    (0.49)                     
          &  7.11    (0.017)            
          &  3.50    ($1.0\xs10^{-5}$)\reject   
          &  3.50    ($1.3\xs10^{-5}$)\reject  \\
          & 14.12    (0.52)                     
          &  2.67    (0.021)            
          &  4.00    ($1.7\xs10^{-5}$)\reject   
          &  4.00    ($3.4\xs10^{-5}$)\reject  \\
\hline % ___________________________________________________________________%|
\crisix   & 14.00    (0.50)              
          &  2.98    ($5.7\xs10^{-4}$)\reject  
          &  5.00    ($1.0\xs10^{-5}$)\reject   
          &  5.00    ($1.0\xs10^{-5}$)\reject  \\
$n=23$    &  7.00    (0.56)              
          &  7.77    (0.0013)           
          &  2.50    ($1.0\xs10^{-5}$)\reject   
          &  2.50    ($1.0\xs10^{-5}$)\reject  \\
$m=108$   &  3.00    (0.61)              
          &  3.71    (0.0018)           
          &  3.00    ($1.4\xs10^{-5}$)\reject   
          &  3.00    ($1.9\xs10^{-5}$)\reject  \\
          &  3.87    (0.69)              
          & 13.54    (0.012)            
          &  7.00    ($1.9\xs10^{-5}$)\reject   
          &  3.50    ($2.9\xs10^{-5}$)\reject  \\
          &  2.50    (0.75)              
          &  2.52    (0.013)            
          &  3.50    ($2.2\xs10^{-5}$)\reject   
          &  3.75    ($3.8\xs10^{-5}$)\reject  \\
\hline % ___________________________________________________________________%|
\criseven & 20.63   (0.083)              
          & \nostat                     
          &  4.00   (\never)\reject         
          & \nostat                    \\
$n=13$    &  5.76   (0.11)               
          & \nostat                     
          &  \nostat                     
          & \nostat                     \\
$m=87$    &  3.82   (0.11)               
          & \nostat                     
          &  \nostat                     
          & \nostat                     \\
          &  2.30   (0.12)               
          & \nostat                     
          &  \nostat                     
          & \nostat                     \\
          &  2.95   (0.12)               
          & \nostat                     
          &  \nostat                     
          & \nostat                     \\
\hline % ___________________________________________________________________%|
\crieight &  2.76   (0.0031)              
          & \nostat                     
          & \nostat         
          & \nostat                \\
$n=8$     &  3.68  (0.064)               
          & \nostat                     
          &  \nostat                     
          & \nostat                     \\
$m=105$   &  3.22   (0.13)               
          & \nostat                     
          &  \nostat                     
          & \nostat                     \\
          & 11.03   (0.31)               
          & \nostat                     
          &  \nostat                     
          & \nostat                     \\
          & 13.80   (0.39)               
          & \nostat                     
          &  \nostat                     
          & \nostat                     \\
\hline % ___________________________________________________________________%|
\end{tabular}
\end{table*}
\addtolength{\tabcolsep}{-0.12cm}  
\renewcommand{\arraystretch}{1.00}
%% __________________________________________________________________________|

\section{Analysis}

Two types of techniques have been mainly applied to search for
periodicity in the impact crater record: the power spectrum
method and those discussed by Yabushita (1991). 
The 28.4 Myr cycle was detected with the former technique
(Alvarez \& Muller 1984).
Techniques of the latter type have been applied,
e.g. by Yabushita (1991) and Grieve \& Pesonen (1996).
Yet it has not been fully realized that these techniques are most 
sensitive to uni--modal phase distributions.
These phases are {\it circular} data: a random sample of 
measurements representing $\phi_{\rm i}$ at $t_{\rm i}$
folded with a period $P$ (i.e.
$\phi_{\rm i} \!=\! {\rm FIX}[t_{\rm i}/P]$,
where ${\rm FIX}$ removes the integer part of $t_{\rm i}/P$).
Several nonparametric methods for detecting both uni-- and 
multi--modal $\phi_{\rm i}$ distributions exist. 
We applied two such methods by Kuiper (1960: the \K) and
Swanepoel \& De Beer (1990: the \SD), and their weighted
versions (Jetsu \& Pelt 1996: the \SHORTWK ~and the {\WSD}s),
which can utilize the additional information in
\wwt ~and \wwd. Our notations are
as in Jetsu \& Pelt (1996).
 
The limit $t\ge5$ eliminates a bias
(\crione, \crithree, \crifour ~and \crisix),
because over 17 \% of $t_{\rm i}$ in Table 1
are below this limit, e.g. these $t_{\rm i}$
have $\phi_{\rm i}\!\leq\!0.05$ for $P\!\ge\!100$. 
A limit in $t$ is unnecessary for \critwo ~and \crifive,
since the criteria in $\sigma_{\rm t}$ and $D$
eliminate most of the $t_{\rm i}\!\le\!5$. 
The criterion in $D$ was applied, 
because earlier studies have
indicated different periodicities for $t_{\rm i}$
of larger and smaller craters (e.g. Yabushita 1991).

The statistical ``null hypothesis'':

$H_0$: {\it ``The $\phi_{\rm i}$ of $t_{\rm i}$ with an
arbitrary period $P$ are randomly distributed between 0 and 1.''}

{\noindent was tested. All tests were performed between
$P_{\rm min}\!=\!2.2$ and $P_{\rm max}\!=\!200$. 
The preassigned significance level 
for rejecting $H_0$ was $\gamma \!=\! 0.001$. 
Two examples of these tests are displayed in Figs. 1 and 2.
Most of the detected $P$ reaching $\gamma \!=\!0.001$ are integers or 
ratios of two integers (see Table 3). 
We call this regularity arising from the bias of assigning integer 
values for $t_{\rm i}$ as the \signal.
Unfortunately, any arbitrary
$P$ can be expressed by a ratio of two integers. 
Two $P_1$ and $P_2$ of the \signal ~will
induce a set of spurious periods
$P'\!=\![P_1^{-1}+k_1(k_2P_2)^{-1}]^{-1}$ 
($k_1\!=\!\pm1,\pm2,...$ and $k_2\!=\!1,2,...$).
Furthermore, another set of spurious periods will be 
$P' \! = \! k_3 P_1 P_2$ ($k_3\!=\!1,2, ...$). 
The \signal ~is strongest
in \crione, since many larger $t_{\rm i}$ end with 5 or 0,
hence \crifour, \crifive ~and \crisix ~were selected.
An arbitrary $P_{\rm min}\!=\!2.2$ was chosen
to avoid the detection of, e.g. $P\!=\!2$, 3/2, and 1.
Note that the order of significance for the period candidates 
in Figs. 1 and 2 is not necessarily the same 
as in Table 3, since these periodograms were derived for 
the overfilling factor $[\Delta \phi]^{-1}\!=\!GM\!=\!10$
(Jetsu \& Pelt 1996: Eq. 12). 
The final values of Table 3 in the vicinity of these
period candidates were determined with $[\Delta \phi]^{-1}\!=\!100$. 
The four nonparametric methods 
are not ``equally sensitive'' to different types of $\phi_{\rm i}$ 
distributions (Jetsu 1996, Jetsu \& Pelt 1996),
which explains the differences in the detected $P$
of Table 3.}

Although uncertainties in some critical
level estimates $Q_{\rm K}$, $Q_{\rm WK}$, $Q_{\rm SD}$ and $Q_{\rm WSD}$
exist, these do not alter the {\it order}
of significance for the detected $P$ (Jetsu \& Pelt 1996).
Thus the \signal ~is the most significant periodicity in the data.
Firstly, the estimate for
the number of independent frequencies in all tests was 
$m \!=\! (f_{\rm max} \! - \! f_{\rm min})/f_0$, 
where $f_{\rm max}=P_{\rm min}^{-1}$,  $f_{\rm min}=P_{\rm max}^{-1}$
and $f_0=(t_{\rm max}-t_{\rm min})^{-1}$. These $m$ estimates
were checked with the empirical correlation 
function $r(k)$ (Jetsu \& Pelt 1996: Eq. 14).
We denote the number of independent frequencies
implied by $r(k)$ with $m'$. Some tests 
had $m' \neq m$. The correct values for the critical level are smaller 
when $m \! > m'$ (i.e. the significance is higher), while the
case is opposite for $m \! < m'$. This problem is only present in \crione ~and
\crifour ~containing less older than younger craters, and thus
the inverse of $t_{\rm max}-t_{\rm min}$ provides
a poor $f_0$ estimate.
Secondly, $Q_{\rm K} \! \ge \! Q_{\rm WK}$ when
the same period is detected with the \SHORTK ~and {\WK}s
(Jetsu \& Pelt 1996: Eq. 30), which is due
to the large scatter in \wwd ~(see Table 2). For example, the sum of the
two largest \wwd ~in \crione($n\!=\!61$)
is 15.25, which disrupts the statistics of the \WK.
However, the values of $Q_{\rm K}$ are reliable when $m\!=\!m'$. 
Thirdly, the statistics of the \WSD ~are quite robust even for a higher
scatter of weights. But the values of $Q_{\rm SD}$ are
$Q_{\rm WSD}$ are uncertain for smaller samples, since the
analytical estimate for the critical parameter $C_{\rm n}$ is
accurate only for larger samples (Jetsu \& Pelt 1996: Eq. 16). 
The \SD ~reveals some bizarre cases ($Q_{\rm SD}\!=\!0\!\equiv$ never),
e.g. for \criseven ~with $P\!=\!4$. No random sample
can contain so many time differences that are multiples of 4.
Even if \criseven ~contained a period that is apparently not 
due to the \signal, such as the 28.4~Myr (Alvarez \& Muller 1984),
it is impossible to decide
whether this period is a multiple of, say 7x4.
In fact, $P\!=\!28.79$ ($Q_{\rm K}\!=\!0.13$)
is the sixth most significant period detected
with the \K ~in \criseven.  

Table 2 shows, why no results for \wwt ~were obtained.  
The largest \wwt ~are typically
$\ga \!n/2$, and thus the statistics
of the \WK ~would be unreliable. As for the \WSD, we refer to the
breakdown parameter $R(s)$ in Table 2. If $R(s)$ exceeds
unity, the statistics of the \WSD ~are disrupted (Jetsu 1996: Eq. 5).
Table 2 indicates that
the \SHORTWK ~and {\WSD}s must not be applied with \wwt. 
Because the high scatter in \wwt ~prevents applications of the
\SHORTWK ~and {\WSD}s, we checked, if the removal of
less accurate $t_{\rm i}$ would significantly alter $w_{\rm max,1}$ or
$R(s)$ for \wwt. But this removal of
less accurate $t_{\rm i}$ did not yield $R(s)\!<\!1$ for \wwt ~even 
if this procedure was carried out
until only 5 values remained in \crione, ..., \criseven. The
same applies to $w_{\rm max,1}$, which remains too close to $n$.

Why was the \signal ~not detected earlier, while the 28.4~Myr cycle was
detected? The answer to the first question is that the $\phi_{\rm i}$ 
distributions
connected to the \signal ~are mostly multi--modal, and can not be detected
with methods sensitive to uni--modal $\phi_{\rm i}$ distributions. 
There are several
answers to the second question. The 28.4 Myr cycle, as well as any other
noninteger period, may be induced by the \signal. For example,
one earlier study revealed mainly multiples of 5 
(Yabushita 1991: e.g. 30x1 =15x2= 10x3 = 6x5).
Since \criseven ~has $Q_{\rm SD}\!=\!0$
for $P\!=\!4$, no Monte Carlo simulation will ever
``succeed'' in producing 
so many exactly equal time differences,
let alone the two additional equal $t$ values in \criseven, 
i.e. the earlier 
significance estimates were not correct (Alvarez \& Muller 1984). 
Finally, the scatter of \wwt ~in \criseven ~is so large that 
the case $n\!=\!13$ does not occur, because the sum
of the two largest \wwt ~is 9.0. An analysis
of the superimposed gaussians of these $t_{\rm i}$ detects
periods from the highest peaks with large \wwt, while the smaller
\wwt ~do not influence the result (e.g. $w_{\rm min}\!=\!0.01$).

The small sample \crieight ~could only be reliably analysed with the \K.
No signs of the 26~Myr cycle were detected (Raup \& Sepkoski 1984, 1986), nor
any period with $Q_{\rm K}\!\le\!\gamma\!=\!0.001$.
The best periods do not betray any trivial signature of the \signal,
although half of the $t_{\rm i}$ are integers. If the most significant
$P\!=\!2.76$ ($Q_{\rm K}\!=\!0.0031$) represents real periodicity,
Raup \& Sepkoski (1984, 1986) have
identified about every tenth mass extinction event. 

\section{Conclusions}

A few topics must be emphasized to avoid misunderstanding.
({\sc{i}}) We did not assume that the integer $t_{\rm i}$
cause the detected periodicities. On the contrary, 
the periodicities were uniquely detected with non--parametric statistical
methods (i.e. model independent).
{\it We simply tested the ``null hypothesis'' ($H_0$) that the 
impact crater ages represent a random sample of circular data.} 
The \signal ~reaches $\gamma \!=\!0.001$ in all
subsamples \crione, ..., \criseven, the critical levels of the \SHORTSD ~and
{\WSD}s being extremely high. The analytical statistics of our methods are
robust, i.e. Monte Carlo or other computational techniques are unnecessary.
({\sc{ii}}) The \signal ~induces irregular
multi--modal $\phi_{\rm i}$ distributions,
which were not detected earlier with methods sensitive to uni--modal
$\phi_{\rm i}$ distributions. For example, the power spectrum method is 
most sensitive to sinusoidal variations.
If the $\phi_{\rm i}$ distribution for the ``correct'' $P'$ 
were exactly bi--modal, the peaks of the power spectrum would be 
at $P'$, $P'/2$, ... But the power spectrum 
method is quite insensitive to more irregular $\phi_{\rm i}$ distributions.
That the \signal ~was not detected earlier, over twelve years after the
study by Alvarez \& Muller (1984), is a direct consequence
of favouring methods sensitive to uni--modal 
$\phi_{\rm i}$ distributions. 
Why should the $\phi_{\rm i}$ 
distributions connected to the possible periodicity in terrestrial
impact cratering rate or mass extinctions of species actually be uni--modal
or of any regular shape?
({\sc{iii}}) It may well be that the large $\sigma_{\rm t_i}$
prevent detection of periodicity (e.g. Heisler \& Tremaine 1989).
In that case our study is simply an exercise of statistics 
providing one new argument against real periodicity. 
However, considering the prevailing theories based on assuming 
the presence of real periodicities (see e.g. Rampino \& Haggerty 1996:
``Shiva Hypothesis''), it was a high time to perform this exercise. 
({\sc{iv}}) The \signal ~has most probably induced those spurious
periods with more or less unimodal $\phi_{\rm i}$ distributions, 
which were detected in several earlier studies (e.g. Alvarez \& Muller 1984,
Yabushita 1991). In any case, the \signal ~is clearly the most significant
periodicity in the impact crater record.
({\sc{v}}) We do not argue that the major comet or asteroid impacts
and the mass extinctions of species are uncorrelated, but emphasize
that the \signal ~dominates the time distribution of the former events.

Our conclusions are simple.
The epochs of mass extinction events of 
species may follow a possibly  ``nonhuman'' cycle of 2.76~Myr, 
but the currently available impact crater
data definitely reveals the embarrassing \signal.
The fellow scientists have unconsciously offered a helping hand to the Nemesis
(e.g. Davis et al. 1984) or ``galactic carrousel'' 
(e.g. Rampino \& Haggerty 1996). The arduous task
for the future geological research is to determine
more accurate (preferably noninteger)
revised ages for impact craters to eliminate the \signal, which may then
lead to a detection of real periodicity.
Over a decade has elapsed in redetecting the 
regularities of our own integer number system and then interpreting them as
periodicity in the ages of impact craters.

\vspace{-0.2cm}
%% __________________________________________________________________________|
\acknowledgements
%% _______________________________________ Do not leave a blank line here!___|
This work was partly supported by the EC Human
Capital and Mobility (Networks) project ``Late type stars: activity,
magnetism, turbulence'' No. ERBCHRXCT940483.
%% __________________________________________________________________________|
%
%                           REFERENCES
%                           ==========
{}
%% __________________________________________________________________________|

\vspace{-0.2cm}
\begin{thebibliography}{}
%% __________________________________________________________________________|

\bibitem{} Alvarez W., Muller R.A. 1984, \nat 308, 718

\bibitem{} Davis M., Hut P., Muller R.A. 1984, \nat 308, 715

\bibitem{} Grewing M., Lequeux J., Pottasch S.R. 1992, \ana 266, E1-E2

\bibitem{} Grieve R.A.F., Pesonen L.J. 1996, Earth, Moon and Planets 72, 357
    
\bibitem{} Heisler J., Tremaine S. 1989, Icarus 77, 213

\bibitem{} Jetsu L., Pelt J. 1996, \anas 118, 587
   
\bibitem{} Jetsu L. 1996, \ana  314, 153

\bibitem{} Kuiper N.H. 1960,
    Proc. Koningkl. Nederl. Akad. Van Wettenschappen, Series A, 63, 38

\bibitem{} Matsumoto M., Kubotani H. 1996, \mnras 282, 1407

\bibitem{} Rampino M.R., Haggerty B.M. 1996, Earth, Moon and Planets 72, 441

\bibitem{} Raup D.M., Sepkoski J.J. 1984, Proc. natn. Acad. Sci. USA 81, 801

\bibitem{} Raup D.M., Sepkoski J.J. 1986, Sci 231, 833

\bibitem{} Swanepoel J.W.H., De Beer C.F. 1990, \apj 350, 754

\bibitem{} Whitmire D.P., Jackson A.A. 1984, \nat 308, 713

\bibitem{} Yabushita S. 1991, \mnras 250, 481
 
\end{thebibliography}
\end{document}

% __________________________________________________________________________|
\addtocounter{table}{+0}               
%
% Notations in text
% =================
% __________________________________________________________________________|
\addtolength{\tabcolsep}{-0.16cm}   
\renewcommand{\arraystretch}{0.20}
\begin{table}
\caption[]{The ages ($t$) of the impact craters of a diameter ($D$)
with an error estimate ($\sigma_{\rm t}$) for $t$. The locations and the
values of
$\sigma_{\rm t}$ and $D$ for the combined events are given in columns 5,
7 and 8, while ``Y'' denotes the samples 
\crione, ..., \crisix.}
\begin{tabular}{rllcccclcccccc}
\hline
 &      &         &              &       &          &  &  &     & & & & &  \\%|
$i$~ & Crater & Country & $t$[Myr]   & $\sigma_{\rm t_1}/\sigma_{\rm t_2}$   %|
& $D$[km]         & $D_1/D_2$    & Location & \crione & \critwo & \crithree  %|
& \crifour & \crifive & \crisix \\                                           %|
 &      &         &              &       &          &  &  &     & & & & & \\ %|
 &      &         &              &       &          &  &  &     & & & & & \\ %|
\hline % _______________________________ first part of SURVEY1.TEX _________ %|
 &      &         &              &       &          &  &  &     & & & & & \\ %|
  1 & 
Kaalij\"arvi         & Estonia        & $0.004\lj0.001$   &             &
0.11                 &                &                                 &
   &   &   &   &   &   \\
  2 & 
Wabar                & Saudi Arabia   & $0.006\lj0.002$   &             &
0.097                &                &                                 &
   &   &   &   &   &   \\
  3 & 
Boxhole              & Australia      & $0.03\lj0.0005$   &             &
0.17                 &                &                                 &
   &   &   &   &   &   \\
  4 & 
Barringer            & U.S.A.         & $0.049\lj0.003$   &             &
1.186                &                &                                 &
   &   &   &   &   &   \\
  5 & 
Lonar                & India          & $0.052\lj0.006$   &             &
1.83                 &                &                                 &
   &   &   &   &   &   \\
  6 & 
Pretoria Saltpan     & South Africa   & $0.22\lj0.052$    &             &
1.13                 &                &                                 &
   &   &   &   &   &   \\
  7 & 
Zhamanshin           & Kazakhstan     & $0.9\lj0.1$       &             &
13.5                 &                &                                 &
   & Y &   &   & Y &   \\
  8 & 
Bosumtwi             & Ghana          & $1.03\lj0.02$     &             &
10.5                 &                &                                 &
   & Y &   &   & Y &   \\
  9 & 
New Quebec           & Canada         & $1.4\lj0.1$       &             &
3.44                 &                &                                 &
   &   &   &   &   &   \\
 10 & 
Tenoumer             & Mauritania     & $2.5\lj0.5$       &             &
1.9                  &                &                                 &
   &   &   &   &   &   \\
 11 & 
Aouelloul            & Mauritania     & $3.1\lj0.3$       &             &
0.39                 &                &                                 &
   &   &   &   &   &   \\
 12 & 
El'gygytgyn          & Russia         & $3.5\lj0.5$       &             &
18                   &                &                                 &
   & Y &   &   & Y &   \\
 13 & 
Roter Kamm           & Namibia        & $3.7\lj0.3$       &             &
2.5                  &                &                                 &
   &   &   &   &   &   \\
 14 & 
Bigach               & Kazakhstan     & $6\lj3$           &             &
7                    &                &                                 &
 Y & Y & Y & Y & Y & Y \\
 15 & 
Shunak               & Kazakhstan     & $12\lj5$          &             &
3.1                  &                &                                 &
 Y &   & Y & Y &   & Y \\
 16 & 
Ries/Steinheim       & Germany        & $15\lj0.7$        & (1/1)       &
24.3                 & (24/3.8)       & N48.9E10.6/N48.0E10.1           &
 Y & Y & Y &   &   &   \\
 17 & 
Haughton             & Canada         & $23\lj1$          &             &
24                   &                &                                 &
 Y & Y & Y & Y & Y & Y \\
 18 & 
Logancha             & Russia         & $25\lj20$         &             &
20                   &                &                                 &
 Y & Y & Y &   &   &   \\
 19 & 
Popigai              & Russia         & $35\lj5$          &             &
100                  &                &                                 &
 Y & Y & Y &   &   &   \\
 20 & 
Chesapeake Bay       & U.S.A.         & $35.5\lj0.6$      &             &
85                   &                &                                 &
 Y & Y & Y & Y & Y & Y \\
 21 & 
Wanapitei            & Canada         & $37\lj2$          &             &
7.5                  &                &                                 &
 Y & Y & Y & Y & Y & Y \\
 22 & 
Mistastin            & Canada         & $38\lj4$          &             &
28                   &                &                                 &
 Y & Y & Y & Y & Y & Y \\
 23 & 
Logoisk              & Belarus        & $40\lj5$          &             &
17                   &                &                                 &
 Y & Y & Y &   &   &   \\
 24 & 
Chiyli               & Kazakhstan     & $46\lj7$          &             &
5.5                  &                &                                 &
 Y & Y & Y & Y & Y & Y \\
 25 & 
Kamensk/Gusev        & Russia         & $49\lj0.14$       & (0.2/0.2)   &
25.2                 & (25/3.5)       & N48.3E40.2/N48.4E40.2           &
 Y & Y & Y & Y & Y & Y \\
 26 & 
Montagnais           & Canada         & $50.5\lj0.76$     &             &
45                   &                &                                 &
 Y & Y & Y & Y & Y & Y \\
 27 & 
Ragozinka            & Russia         & $55\lj5$          &             &
9                    &                &                                 &
 Y & Y & Y &   &   &   \\
 28 & 
Marquez              & U.S.A.         & $58\lj2$          &             &
13                   &                &                                 &
 Y & Y & Y & Y & Y & Y \\
 29 & 
Chicxulub            & Mexico         & $64.98\lj0.05$    &             &
170                  &                &                                 &
 Y & Y & Y & Y & Y & Y \\
 30 & 
Kara/Ust-Kara        & Russia         & $73\lj2.1$        & (3/3)       &
69.6                 & (65/25)        & N69.2E65.0/N69.3E65.3           &
 Y & Y & Y & Y & Y & Y \\
 31 & 
Manson               & U.S.A.         & $73.8\lj0.3$      &             &
35                   &                &                                 &
 Y & Y & Y & Y & Y & Y \\
 32 & 
Lappaj\"arvi         & Finland        & $77.3\lj0.4$      &             &
23                   &                &                                 &
 Y & Y & Y & Y & Y & Y \\
 33 & 
Boltysh              & Ukraine        & $88\lj3$          &             &
24                   &                &                                 &
 Y & Y & Y & Y & Y & Y \\
 34 & 
Dellen               & Sweden         & $89\lj2.7$        &             &
19                   &                &                                 &
 Y & Y & Y & Y & Y & Y \\
 35 & 
Steen River          & Canada         & $95\lj7$          &             &
25                   &                &                                 &
 Y & Y & Y &   &   &   \\
 36 & 
Deep Bay/West Hawk   & Canada         & $100\lj35$        & (50/50)     &
13.2                 & (13/2.44)      & N56.4W103.0/N49.8W95.2          &
 Y &   &   &   &   &   \\
 37 & 
Carswell/Zapadnaya   & Canada/Ukraine & $115\lj7.1$       & (10/10)     &
39.2                 & (39/4)         & N58.4W109.5/N49.7E29.0          &
 Y & Y & Y &   &   &   \\
 38 & 
Zeleny Gai           & Ukraine        & $120\lj20$        &             &
2.5                  &                &                                 &
 Y &   & Y &   &   &   \\
 39 & 
Mien                 & Sweden         & $121\lj2.3$       &             &
9                    &                &                                 &
 Y & Y & Y & Y & Y & Y \\
 40 & 
Tookoonooka          & Australia      & $128\lj5$         &             &
55                   &                &                                 &
 Y & Y & Y & Y & Y & Y \\
 41 & 
Romistrovka          & Ukraine        & $140\lj20$        &             &
2.7                  &                &                                 &
 Y &   & Y &   &   &   \\
 42 & 
Gosses Bluff         & Australia      & $142.5\lj0.5$     &             &
22                   &                &                                 &
 Y & Y & Y & Y & Y & Y \\
 43 & 
Mj{\o}lnir           & Norway         & $143\lj20$        &             &
40                   &                &                                 &
 Y & Y & Y & Y & Y & Y \\
 44 & 
Liverpool            & Australia      & $150\lj70$        &             &
1.6                  &                &                                 &
 Y &   &   &   &   &   \\
 45 & 
Vepriai              & Lithuania      & $160\lj30$        &             &
8                    &                &                                 &
 Y &   &   &   &   &   \\
 46 & 
Puchezh-Katunki      & Russia         & $175\lj3$         &             &
80                   &                &                                 &
 Y & Y & Y &   &   &   \\
 47 & 
Rochechouart         & France         & $186\lj8$         &             &
23                   &                &                                 &
 Y & Y & Y & Y & Y & Y \\
 48 & 
Wells Creek/Red Wing & U.S.A.         & $200\lj18$        & (100/25)    &
15                   & (12/9)         & N36.4W87.7/N47.6W103.6          &
 Y & Y & Y &   &   &   \\
 49 & 
Manicouagan          & Canada         & $214\lj1$         &             &
100                  &                &                                 &
 Y & Y & Y & Y & Y & Y \\
 50 & 
Obolon'              & Ukraine        & $215\lj25$        &             &
15                   &                &                                 &
 Y &   &   &   &   &   \\
 51 & 
Saint Martin         & Canada         & $220\lj32$        &             &
40                   &                &                                 &
 Y &   &   &   &   &   \\
 52 & 
Araguainha Dome      & Brazil         & $247\lj5.5$       &             &
40                   &                &                                 &
 Y & Y & Y & Y & Y & Y \\
 53 & 
Kursk                & Russia         & $250\lj80$        &             &
5.5                  &                &                                 &
 Y &   &   &   &   &   \\
 54 & 
Clearwater West/East & Canada         & $290\lj14$        & (20/20)     &
44.4                 & (36/26)        & N56.2W74.5/N56.1W74.1           &
 Y &   & Y &   &   &   \\
 55 & 
Dobele               & Latvia         & $300\lj35$        &             &
4.5                  &                &                                 &
 Y &   &   &   &   &   \\
 56 & 
Crooked Creek        & U.S.A.         & $320\lj80$        &             &
7                    &                &                                 &
 Y &   &   &   &   &   \\
 57 & 
Charlevoix           & Canada         & $357\lj15$        &             &
54                   &                &                                 &
 Y &   &   & Y &   &   \\
 58 & 
Flynn Creek          & U.S.A.         & $360\lj20$        &             &
3.55                 &                &                                 &
 Y &   &   &   &   &   \\
 59 & 
Siljan               & Sweden         & $368\lj1.1$       &             &
52                   &                &                                 &
 Y &   &   & Y &   &   \\
 60 & 
Kaluga               & Russia         & $380\lj10$        &             &
15                   &                &                                 &
 Y &   &   &   &   &   \\
 61 & 
Ilyinets             & Ukraine        & $395\lj5$         &             &
4.5                  &                &                                 &
 Y &   &   &   &   &   \\
 62 & 
La Moinerie          & Canada         & $400\lj50$        &             &
8                    &                &                                 &
 Y &   &   &   &   &   \\
 63 & 
Couture              & Canada         & $430\lj25$        &             &
8                    &                &                                 &
 Y &   &   &   &   &   \\
 64 & 
Calvin/Brent         & U.S.A./Canada  & $450\lj7.1$       & (10/30)     &
9.3                  & (8.5/3.8)      & N41.8W86.0/N46.1W78.5           &
 Y &   &   &   &   &   \\
 65 & 
Pilot                & Canada         & $455\lj2$         &             &
6                    &                &                                 &
 Y &   &   &   &   &   \\
 66 & 
Ames                 & U.S.A          & $470\lj30$        &             &
16                   &                &                                 &
 Y &   &   &   &   &   \\
 67 & 
Gardnos              & Norway         & $500\lj10$        &             &
5                    &                &                                 &
 Y &   &   &   &   &   \\
 68 & 
Holleford            & Canada         & $550\lj100$       &             &
2.35                 &                &                                 &
 Y &   &   &   &   &   \\
 69 & 
Mizarai              & Lithuania      & $570\lj50$        &             &
5                    &                &                                 &
 Y &   &   &   &   &   \\
 70 & 
J\"anisj\"arvi       & Russia         & $698\lj22$        &             &
14                   &                &                                 &
 Y &   &   & Y &   &   \\
 71 & 
Morokweng            & South Africa   & $1400\lj200$      &             &
70                   &                &                                 &
 Y &   &   &   &   &   \\
 72 & 
Teague               & Australia      & $1630\lj5$        &             &
30                   &                &                                 &
 Y &   &   &   &   &   \\
 73 & 
Sudbury              & Canada         & $1850\lj3$        &             &
250                  &                &                                 &
 Y &   &   &   &   &   \\
 74 & 
Vredefort            & South Africa   & $2018\lj14$       &             &
300                  &                &                                 &
 Y &   &   & Y &   &   \\
\hline % ___________________________________________________________________ %|
\end{tabular}
\addtolength{\tabcolsep}{0.16cm}  
\renewcommand{\arraystretch}{1.00}
\end{table} % _______________________________________________________________|